\documentclass{aa}

\usepackage{graphicx}
\usepackage[table,xcdraw]{xcolor}
\usepackage{hyperref}
\hypersetup{colorlinks=true, citecolor=blue}
\usepackage{enumitem}
\graphicspath{{figs/}}
\usepackage{float}

\newcommand{\fz}{${\rm d}^2N/{\rm d}z{\rm d}W$}
\newcommand{\dndx}{${\rm d}N/{\rm d}X$}
\newcommand{\dndz}{${\rm d}N/{\rm d}z$}
\newcommand*\mean[1]{\overline{#1}}

\newcommand{\Waver}{\mean W_r(R)}

\newcommand{\kms}{km~s$^{-1}$}

\newcommand{\mgii}{\ion{Mg}{ii}}
\newcommand{\hi}{\ion{H}{i}}

\newcommand{\oi}{\ion{O}{i}}

\newcommand{\hidden}[1]{}

\begin{document}

\title{
  How far have metals reached? 
  Reconciling statistical constraints
  and enrichment models at reionization
}
\author{
  Sebastian Lopez\inst{1,}\thanks{\email{slopez@das.uchile.cl}}
  \and
  Jens-Kristian Krogager\inst{2,3}
  }

\institute{Departamento de Astronom\'ia, Universidad de Chile,
  Casilla 36-D, Santiago, Chile. 
  \and
  French-Chilean Laboratory for Astronomy, IRL 3386, CNRS and U. de Chile, Casilla 36-D, Santiago, Chile
  \and
  Centre de Recherche Astrophysique de Lyon, Universit{\'e} de Lyon 1, UMR5574, 69230 Saint-Genis-Laval, France
}
\titlerunning{Evolution of metal-enriched bubbles}
\authorrunning{Lopez \& Krogager}

\abstract{ { The incidence of quasar absorption systems and the space
    density of their galaxies are proportional, with the
    proportionality factor given by the mean absorbing cross section.
    In this paper, we use redshift parametrizations of these two
    statistics to predict the cosmic evolution of an equivalent-width
    ($W_r$) radial profile model, tailored for the low-ionization
    species \mgii\ and \oi.  Our model provides an excellent match to
    well-sampled, low-redshift
    \mgii\ equivalent-width and impact-parameter pairs from the
    literature. 
    We then focus on the evolution of various quantities
    between the reionization and cosmic noon eras.  We find that the
    extent of \mgii, and hence the amount of cool ($T\sim 10^4$ K),
    enriched gas in the average halo, decreases continuously with
    cosmic time, suggesting that the expected growth of metal-enriched
    bubbles before reionization experienced a turnover in its
    low-ionization phase at around $z \approx 6$--$8$.  This effect is
    more pronounced in $W_r^{2796}\lesssim 0.3$ \AA\ systems
    (outermost layers of the model) and, in general, affects \oi\ more
    than \mgii,} probably owing to the onset of photoionization by the
  UV background.  The line density of $W_r^{2796}\gtrsim 1$ \AA\ 
  systems (model inner layers) continuously increases in synchrony
  with the star-formation rate density until it reaches a peak at
  cosmic noon. In contrast, the line density of $W_r^{2796}\lesssim
  0.3$ \AA\ systems remains constant or decreases over the same
  period.  (3) At the end of reionization, the filling factor is low
  enough so that the winds have not yet reached neighboring
  halos. This implies that the halos are self-enriched, as suggested
  by semi-analytic models, through a process combined with the
  constant replenishment of the intergalactic medium. We discuss how
  these statistical predictions can be reconciled with early metal
  enrichment models and argue that they offer a practical comparison
  point for future analyses of quasar absorption lines at $z>6$.  }

  \maketitle

\section{Introduction} 
\label{introduction}
The James Webb Space Telescope has revealed a population of very high
redshift galaxies whose space densities and luminosities increase
dramatically with cosmic time towards the end of
reionization~\citep[e.g., ][]{Willott2024,Whitler2025}. While there is
growing evidence that these galaxies host the massive stars that
ionized the surrounding intergalactic medium~\citep[IGM; e.g.,
][]{Furlanetto2004}, it is possible that these same stars also
enriched the IGM with metals~\citep[e.g.,
][]{Shapiro1994,Miralda-Escude1998b,Kashino2023}. It has long been
predicted that the Universe experienced early and widespread metal
enrichment via supernova explosions and
superwinds~\citep{Ferrara2000,Madau2001,Madau2014,Heckman2000}. This
has finally begun to be observed in the early Universe~\citep[e.g.,
][]{Carniani2024} and is consistent with galaxy luminosity
functions~\citep{Ferrara2024}.  Metals in the earliest galaxies
produced by the first supernova explosions are thought to have resided
in low-ionization stages~\citep{Furlanetto2003}, inside regions later
ionized during reionization~\citep{Bolton2013}.  Thus, the
reionization of the Universe is intimately linked to its early metal
enrichment.

  In this paper, we constrain the extent of low-ionization metals
  after reionization, as detected in absorption toward background
  quasars.  We adopt the standard geometric assumption that 
    the probability of intercepting metal-enriched halos
    is proportional to both their number density and their cross
    section. 
  Pioneering studies at
  $z\la 2$ related this probability to the observed incidence of
  low-ionization absorbers to infer their cross sections.  Sizes of
  approximately $10$--$15$ proper kiloparsecs (pkpc) were found for
  damped Lyman-alpha (Ly$\alpha$) disks~\citep[][]{Wolfe1986} and
  about $40$--$60$ pkpc for very strong ($W_r^{2796}\ge 1$ \AA)
  \mgii\ systems~\citep[][]{Lanzetta1991,Steidel1995,Churchill1996},
  using halo-size-luminosity scaling relations and covering fractions
  inferred from direct galaxy associations.  These early applications
  laid the groundwork for later circumgalactic medium (CGM) models and
  absorber–halo scaling relations~\citep[e.g.,
  ][]{Chen2000,Fynbo2008,Kacprzak2008,Krogager2020a}.  Subsequent
  studies using larger and deeper samples of \mgii\ systems showed
  that halo size scales only weakly with galaxy luminosity and has
  undergone little evolution up to $z\sim
  1.5$~\citep{Churchill2000,Nielsen2013b,Chen2010a,Chen2010b}. Later,
  \citep[][]{Tinker2008,Tinker2010} applied a halo occupation model to
  constrain the extent and halo-mass dependence of \mgii, suggesting
  an evolving relationship between halo size and absorption cross
  section up to $z\approx 2$.  With the advent of near-infrared quasar
  surveys, exploration of \mgii\ beyond redshift $z=2$
  began~\citep{Matejek2012,Codoreanu2017,Chen2017,Bosman2017,Becker2019},
  revealing significant redshift evolution. 
  Between 
  $z\approx 2$ 
  and 
  $z\approx 6$,  
    the comoving line density of very strong
  \mgii\ systems is shown to decrease by a factor of $\sim
  5$~\citep{Chen2017} while their cross section increases by a factor
  of $\sim 3$~\citep{Seyffert2013,Codoreanu2017}. 
Here we present a simple formalism that, unlike previous work limited to estimating average cross sections {  for a single
  equivalent width threshold}, computes $W_r$ radial
profiles and volume filling factors as a function of redshift. 
  We combine the most recent measurements of the
  high-redshift luminosity function of star-forming galaxies with our
  own redshift parametrization of the highest-redshift 
  line frequency distribution ($z\la7$) to date. This
  approach, together with the new, deeper 
  data available, allows us to study the evolution of  low-ionization metal-enriched gas as a
  function of absorption strength.  We begin by 
  presenting the formalism of our model in
  \S~\ref{sect_formalism} and its implementation at low and high
  redshifts in \S~\ref{sect_implementation}. In \S~\ref{sect_results}, we present the
  results on the predictions of
  equivalent-width radial profiles, radial extent, and filling
  factors. We discuss our findings in \S~\ref{sect_discussion} and
  summarize our conclusions in \S~\ref{sect_conclusions}.  
Throughout the paper, we use a $\Lambda$CDM cosmology with the
following cosmological parameters: $H_0=70$
\,km\,s$^{-1}$\,Mpc$^{-1}$, $\Omega_M=0.3$, and $\Omega_{\Lambda}
=0.7$.

\section{Formalism}
\label{sect_formalism}

We assume that all UV-bright galaxies are surrounded by enriched
halos, such that if the halo of a galaxy of luminosity, $L$, and
redshift, $z$, is crossed by the line of sight to a background quasar at
the projected distance, $\rho$, from the galaxy, a rest-frame equivalent
width, $W_r$, is expected at $z$. We further assume that a function 
\begin{equation}
W_r=W_r(\rho, L, z) 
\label{eq_W}  
\end{equation}
exists, which decreases monotonically with $\rho$. This condition
{  is motivated by the well-known  
$W_r$-$\rho$ anti-correlation of \mgii, identified by various
  techniques up  to  
  $\rho\approx100$ kpc at
  $z\sim
  1$~\citep[e.g.,][]{Chen2010a,Bordoloi2011,Nielsen2013b,Rubin2018a,Lopez2018,Dutta2020,Lundgren2021,Cherrey2025,Berg2025,Das2025}.    
In the context of our model, it also 
}
implies
that the inverse function $\rho = \rho(W_r)$ is well defined at a
given $z$. Eq.~\ref{eq_W} assumes a single population of homogeneous and spherically symmetric halos. It does not address galaxy
orientation or complex feedback effects in a multiphase
CGM~\citep[e.g., ][]{Tumlinson2017,Guo2023}. In a
pencil-beam survey, the equivalent width distribution 
$f(z,W)\equiv$~\fz\ 
is defined as the number of systems, $N$,
with $W_r$ between $W$ and $W+dW$ per
unit redshift; 
i.e., integration over $W$ gives the line density,
\dndz. Conversely, \dndz\ is proportional to the
probability of intersecting a halo. Hence, under the above
assumptions, we obtain~\citep[e.g., ][]{Hogg1999}
\begin{equation}
  f(z,W){\rm d}W=n_c~{\rm d}\sigma~\frac{c}{H_0}\frac{{\rm d}X}{{\rm d}z},
\label{eq_cosmo}  
\end{equation}
where $n_c$ is the comoving space density of galaxies and  $d\sigma$ is
the physical cross section, in Mpc$^2$, 
where absorbers of $W_r$
occur\footnote{The ``differential absorption path length'' 
term~\citep{Bahcall1969}, 
d$X/$d$z\equiv(1+z)^2/\sqrt{\Omega_M (1 + z)^3 +  \Omega_\Lambda}$,
accounts for the cosmological change in the probability of 
intersection.}. 
In integral form, 
\begin{equation}
  \int^{\infty}_{\mean W_r} f(z,W){\rm d}W = \frac{c}{H_0}\frac{{\rm d}X}{{\rm d}z}~\int^{L_{\rm
      max}}_{L_{\rm min}}\phi(z,L)~\sigma(L,z){\rm d}L~.
\label{eq_hits}
\end{equation}  
Here, $\phi(z,L)$ is the luminosity function of galaxies and
${\mean W_r}$ is the {  mean rest frame} equivalent width {  per
  galaxy.} 
Note that $f(z)$, $\phi(z)$, and therefore also 
$\sigma(z)$, can all be functions of 
redshift, independently of Hubble flow.
Eq.~\ref{eq_hits} allows us to numerically obtain $\sigma(z)$.

\begin{figure}
  \centering
  \includegraphics[width=1\columnwidth]{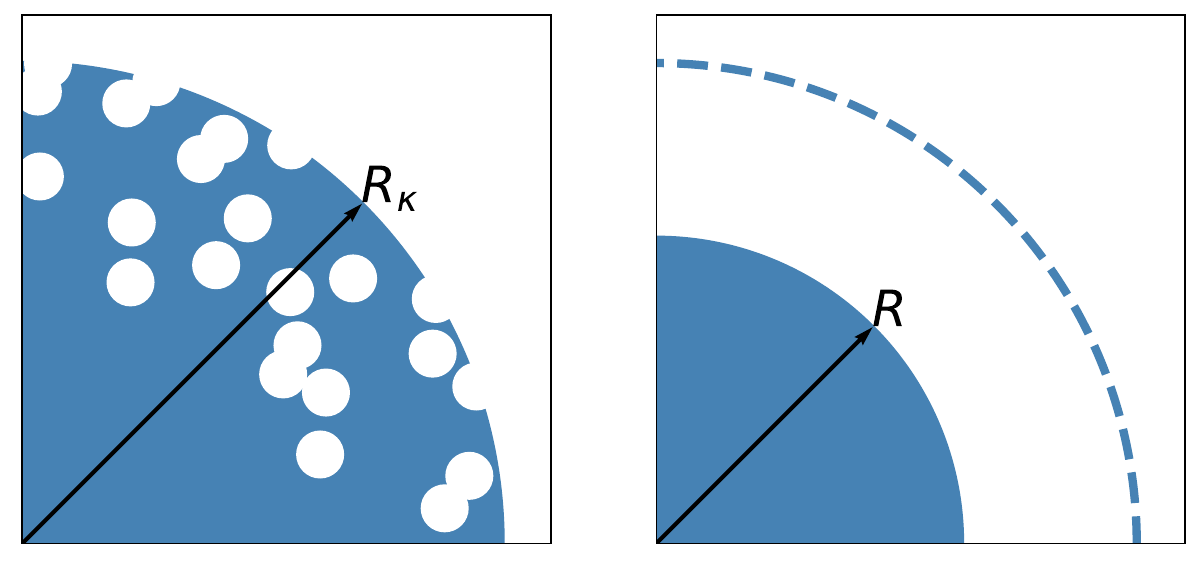}
  \caption{ {  Sky-plane} representation of a single equivalent
    width layer illustrating the cases of non-unity (left) and unity covering
    fraction.
    \label{fig_cheese}
    }
\end{figure}

Essential to our goals, Eq.~\ref{eq_hits} is agnostic regarding how
$\sigma$ is actually distributed on the sky plane. Therefore,
obtaining a linear scale from $\sigma$ requires assumptions about
its distribution, hence the importance of assuming a decreasing
profile.  Observationally, $f(z,W)$ is corrected not only for survey
incompleteness but also for the redshift path that does not give rise to
absorption, due to non-unity covering fraction~\cite[$\kappa$;
  e.g.,][]{Chen2010a,Nielsen2015,Lan2018}.  Therefore, any area
inferred from Eq.~\ref{eq_hits} is already factorized by $\kappa$.
{  Here, we define $\kappa=\kappa(\rho)$ as the fraction of area
  giving rise to absorption at a given $\rho$, i.e., above a given
  $W_r$ in our layered model}.  Throughout this work we assume a
circular layer structure.  This makes ${\mean W_r}$ a
twofold average, including the spatial average over azimuthal angles.
But the concept of radial profile and ``radius'' makes sense only if
$\kappa$ is known from  independent observations.  A simplified
situation for a {  single} $W_r$ is shown in
Fig.~\ref{fig_cheese}.  Under the assumption of non-unity covering
fraction (left-hand panel), we can define a radius $R_\kappa$ such
that it contains all the absorbing footprint, regardless of its
spatial distribution.  Conversely, under the assumption of unity covering
fraction (right-hand panel), the survey yields the same absorbing
footprint {  but} a smaller radius, $R$.  {  Therefore,} the two
radii must follow the relation 
\begin{equation}
\sigma = 2\pi\int_0^R \rho {\rm d}\rho = 2\pi\int_0^{R_{\kappa}} \kappa(\rho)\rho
       {\rm d}\rho~.
       \label{eq_kappa}
\end{equation}

\section{Implementation}
\label{sect_implementation}
We implemented numerical solutions for $\Waver$ using
parametrizations of $f$ and $\phi$. We emphasize that such
parametrizations are supported by completely independent data sets: 
absorption systems and galaxies. For $f(W)$,
we used results {  on the \mgii\ $\lambda 2796$ transition} by
various authors, who provide different 
parametrizations depending on survey quality. The most general
parametrization is a Schechter
function~\citep{Kacprzak2011,Mathes2017,Bosman2017}:
\begin{equation}
  f(z,W_r) = \left(\frac{N^\star}{W^\star}\right)
  \left(\frac{W^{}_r}{W^\star}\right)^{\alpha}
  e^{-\left(\frac{W_r}{W^\star}\right)}~,
  \label{eq_f}
\end{equation}
where $N^\star$ is a normalization factor, $\alpha$ is the weak-end
power-law index, and $W^\star$ is the turnover equivalent width.  { 
  For \mgii, the adopted ``weak'' limit is
  $W_r^{2796}=0.3$ \AA~\citep{Churchill1999}. }  Absorption-line surveys are
quite heterogeneous: usually, $\alpha$ cannot be measured at resolving
power $\mathcal{R}\lesssim 2\,000$; likewise, $W^\star$ is not well
constrained if the redshift path is insufficient to deal with the
low-number statistics at the strong end.  In such cases, an exponential
function is 
used to fit the \fz\ data.  For $\phi(L)$ we used the UV
luminosity function from~\citet{Bouwens2021}, {  also parametrized using
a Schechter function.} This is the latest and most comprehensive
compilation to date, providing independent redshift parametrizations 
for all three Schechter parameters up to $z\sim 10$.

 To solve the right-hand integral in Eq.~\ref{eq_hits} it is usually assumed  that halo ``sizes'' scale with
luminosity~\citep[e.g.,][]{Guillemin1997,Chen1998,Kacprzak2008,Chen2010a}. 
 In this case, we substitute $R$ by $R(L/L^\star)^\beta$ and from
  Eq.~\ref{eq_kappa} we obtain $\sigma(L,z) = \pi
R^2(L/L^\star)^{2\beta}$ (and likewise for $R_\kappa$),   where $R=R({\mean W_r})$ is now the radius corresponding to an $L^{\star}$ galaxy.    The 
  integral then becomes a luminosity-weighted comoving space density
  of galaxies, $\langle n\rangle_L \equiv \int
  \phi(L)(L/L^\star)^{2\beta}dL$.   Assumptions must be made about the
redshift evolution of $\beta$.  A dependence of $W_r$ on luminosity
is expected through star-formation rate (SFR) 
and therefore stellar mass, 
$M_{star}$~\citep{Chen2001,Huang2021,Weng2024}. Since $M_{star}$
correlates strongly with luminosity~\citep[e.g., ][]{Moster2010}, the
parameter $\beta$ breaks down into a geometrical ($\rho$) and a
physical ($W_r$) component; i.e., $\beta$ could be $0.5$ if
$W_r\propto
M_{star}^{0.5}$~\citep{Lan2018,Krogager2020a,Chen2025}. Here, we simply
adopted $\beta=0.35(1+z)^{0.2}$.  {  With these prescriptions, ${\mean
    W_r} = {\mean W_r}(R)$ can be solved by substituting
  Eq.~\ref{eq_kappa} into Eq.~\ref{eq_hits} and evaluating it
  numerically over a running range of equivalent
  widths\footnote{Alternatively, Eq.~\ref{eq_cosmo} leads to a
    non-linear differential equation of the form d$W+Cf(W)^{-1}2\pi
    \kappa(\rho)\rho{\rm d}\rho=0$, where $C>0$ is a constant.}.

\begin{figure}
  \centering
  \includegraphics[width=1\columnwidth]{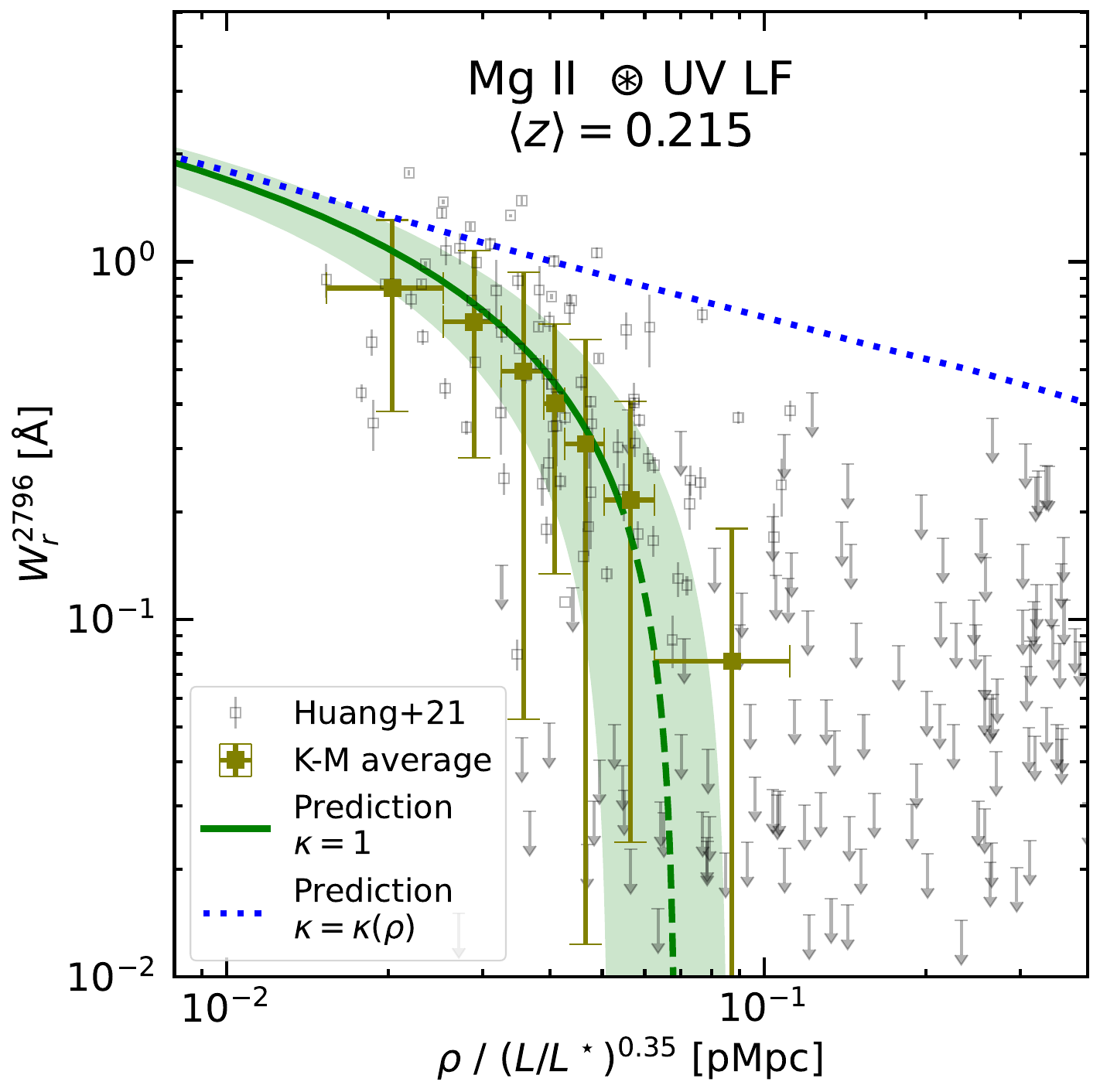}
  \caption{ Rest-frame equivalent width of \mgii\ vs. projected
    separation of the absorbing galaxy.  Black symbols represent direct data 
   from~\citet{Huang2021}.  Olive symbols mark average $W_r$ values
    weighted by the probability distribution function (PDF) of a Kaplan-Meier estimator of
    detections and non-detections in the $\rho$-bin (details in
    the text).  The green line shows the statistical prediction of the $\Waver$
    using Eqs.~\ref{eq_hits} and~\ref{eq_kappa} for the case
    $\kappa=1$.  The function $f(W)$ is taken from~\citet{Zhu2013},   
    extrapolated to $\langle z\rangle=0.215$, while $\phi(z,L)$ is  
    from~\citet{Bouwens2021} with $L_{min}/L^\star=0.03$ and
    $L_{max}/L^\star=7.5$.  The shaded region propagates measurement
    errors of both statistics.  The green dashed portion 
    indicates {  the weak-\mgii\ regime where $f(W)$ is incomplete.}
  The dotted blue line shows the 
    statistical prediction of $\mean W_r(R_\kappa)$, applying a 
    low-redshift extrapolation of $\kappa=\kappa(\rho)$ {  proposed} 
    by~\citet{Schroetter2021} {  to model} an independent data set.
        \label{fig_lowz}
    }
\end{figure}

\subsection{Low redshift}
\label{sect_lowz}

We tested our approach with $W_r$-$\rho$ data obtained
by~\citet{Huang2021} for \mgii\ systems around low-redshift isolated
galaxies (gray symbols in Fig.~\ref{fig_lowz}).  A key property of
this survey is that the galaxy-quasar pairs were selected with no prior
knowledge of the \mgii, thus yielding a realistic estimate of the
covering fraction. The sample comprises 211 galaxy-quasar pairs with
galaxy redshifts and luminosities in the range $(0.0968,0.4838)$ and
$(0.01,7.5)L^\star$, respectively. To compare with the prediction on
$\Waver$ we obtained averages of $W_r$ in radial bins.  To account for
non-detections, we used a survival analysis for censored
data~\citep{Feigelson1985}. In particular, we used the Kaplan-Meier
(K-M) estimator implemented in the `lifelines’  
library~\citep{Davidson-Pilon2019}. 
}
First, the data were binned in
$\rho$ bins of equal size (olive-colored horizontal bars in the
figure). Then a ``time'' average and a 1-$\sigma$ dispersion were
calculated (squares and vertical error bars) by weighting the scores in
each bin by the survival probability density function. Our predicted
$\Waver$ (green line) was computed by solving equations~\ref{eq_hits}
and~\ref{eq_kappa} within the~\citet{Huang2021} luminosity range, at
the survey median redshift $\langle z \rangle=0.2151$. We used $f(z,W)$
by~\citet{Zhu2013}.  This frequency distribution is based on SDSS
$\mathcal{R}\approx 1,800$ spectra (i.e., incomplete below $W_r^{2796}\approx 0.3$ \AA) and is described solely by the exponential part of
Eq.~\ref{eq_f}.  Overall, redshifts $z\lesssim 0.4$ are not covered,
so we resort to those authors' redshift parametrization and
extrapolate $f(z,W)$ to $\langle z \rangle$. Overall, our statistical
prediction matches the binned data well, regardless of bin size
(the root mean square error is $\approx 10$\%, excluding the lowest
$W_r$ bin).  The shaded region indicates the 1-$\sigma$ confidence
interval 
obtained by propagating
the errors of both statistics, though it is dominated by the uncertainties in
$f(z,W)$. We again stress that the prediction is based on
observational data, and the three data sets involved have totally
independent origins.  The dashed green portion of the curve is an
extrapolation to $W_r\lesssim 0.2$ \AA, a regime that is not covered by the
line density data used here.  It is not surprising that
the outermost $W_r$ bin is underpredicted, since higher-resolution 
work~\citep[e.g., ][]{Kacprzak2011} has shown that the    
power-law part of $f(z,W)$ (Eq.~\ref{eq_f}) outnumbers the strong
$W_r$ line density. As a consequence, if the same is valid at low
redshift, a wider $\Waver$ profile and a better fit in this bin would
be expected. There is currently no measured weak line density
statistics at these low redshifts.

The above prediction is for $\kappa=1$.  The dotted blue line shows
 $\mean W_r(R_\kappa)$ 
using $\kappa=\kappa(\rho)$ in Eq. 7 of~\citet{Schroetter2021}, also
extrapolated to $\langle z \rangle$.  Unsurprisingly, this profile
encompasses almost all data points (including upper limits). The few
points above the prediction are concentrated at $\rho\lesssim 70$
kpc. In the present approach, these are classified as statistical
outliers. We speculate that they may correspond to galaxy-scale
processes, e.g., 
outflows~\citep{Weiner2009,Bradshaw2013}, 
streams~\citep{Waterval2025}, and/or random galaxy
orientations that break the spherical symmetry assumed here.
Altogether, we conclude that coupling the observational
line density and luminosity-function data provides an excellent
prediction for the independently observed $W_r$-$\rho$ data  at low
redshift; moreover, the luminosity limits used in Eq.~\ref{eq_hits}
naturally yield the correct normalization of $\Waver$.
Excluding the uncertain weak-\mgii\ statistics, this finding
suggests that most of the strong \mgii\ is indeed associated with UV-luminous galaxies, with the caveat that quiescent and post-starburst
galaxies constitute only a few percent of UV-bright galaxies~\citep[e.g., 
][]{Taylor2023} in any environment, so their
effect~\citep{Lan2018,Chen2025}
could go unnoticed in our model. 
\subsection{Redshift dependence}
\label{sect_highz}

\begin{figure}
  \centering
  \includegraphics[width=1\columnwidth]{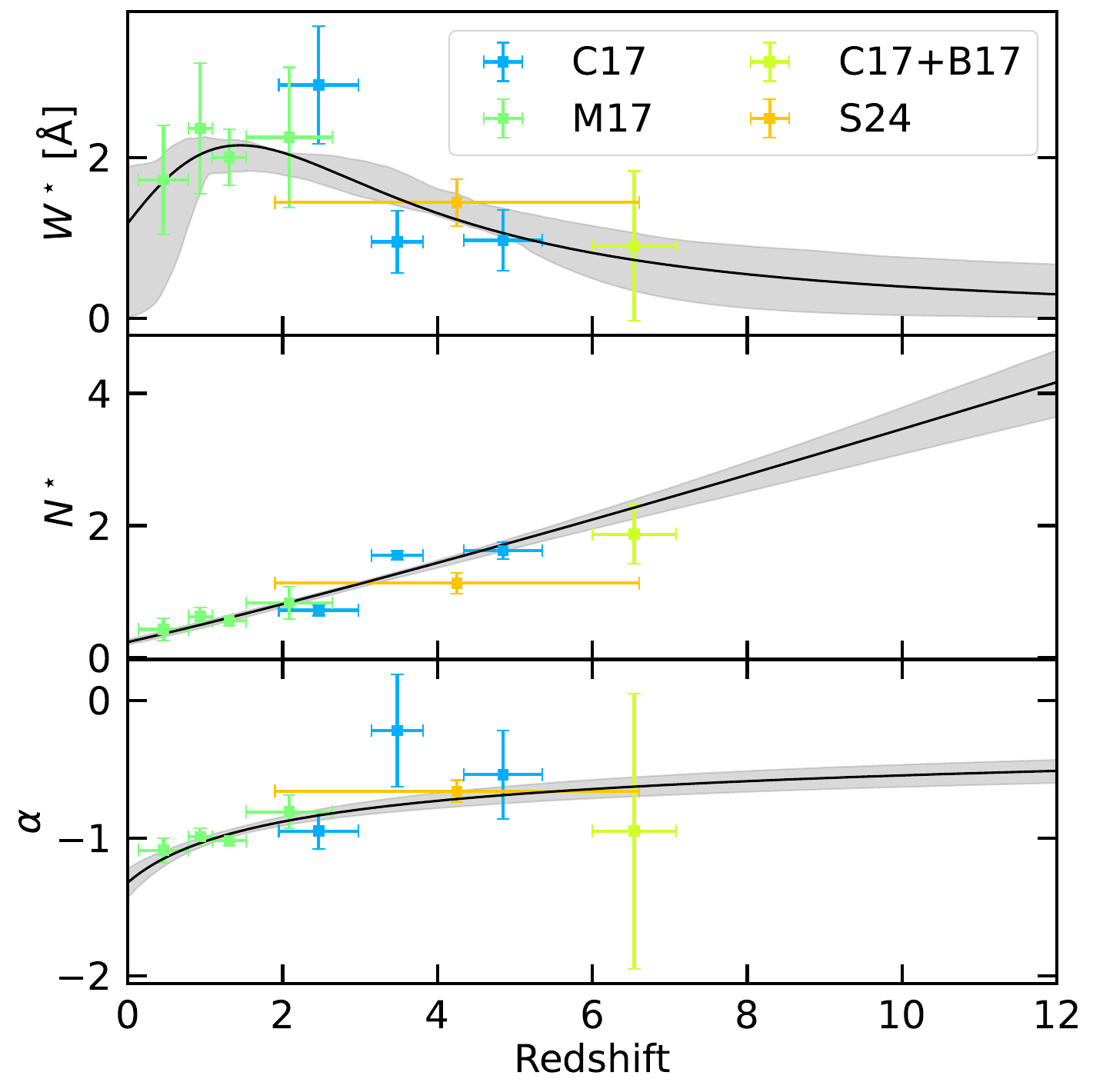}
  \caption{ Redshift evolution of the Schechter parameters that fit the
    \fz\ data from~\citet[][M17]{Mathes2017}, \citet[][C17]{Chen2017}, \citet[][B17]{Bosman2017}, and~\citet[][S24]{Sebastian2024}. The M17 and S24 data points are published, while the C17 
    and {  C17+}B17 data points are obtained by us {  by re-fitting
    the authors' \fz\ data } (see caption of
    Fig.~\ref{fig_schechter}).  The solid curves are parametrizations
    of the form $y=a(1+z)^b /(1+ ((1+z)/c)^d)$ fit to the $W^\star$
    data, and $y=a(1+z)^b$ fit to the $N^\star$ and the $\alpha$
    data. The best-fit parameters are listed in
    Table~\ref{table_fit}. The 1-$\sigma$ bands are computed using
    bootstrapping over the data ($W^\star$) and covariance-based
    bootstrapping over the fit parameters ($N^\star$ and $\alpha$).
  }
  \label{fig_f_params_schechter}
\end{figure}

\begin{figure*}
  \centering
    \includegraphics[width=2\columnwidth]{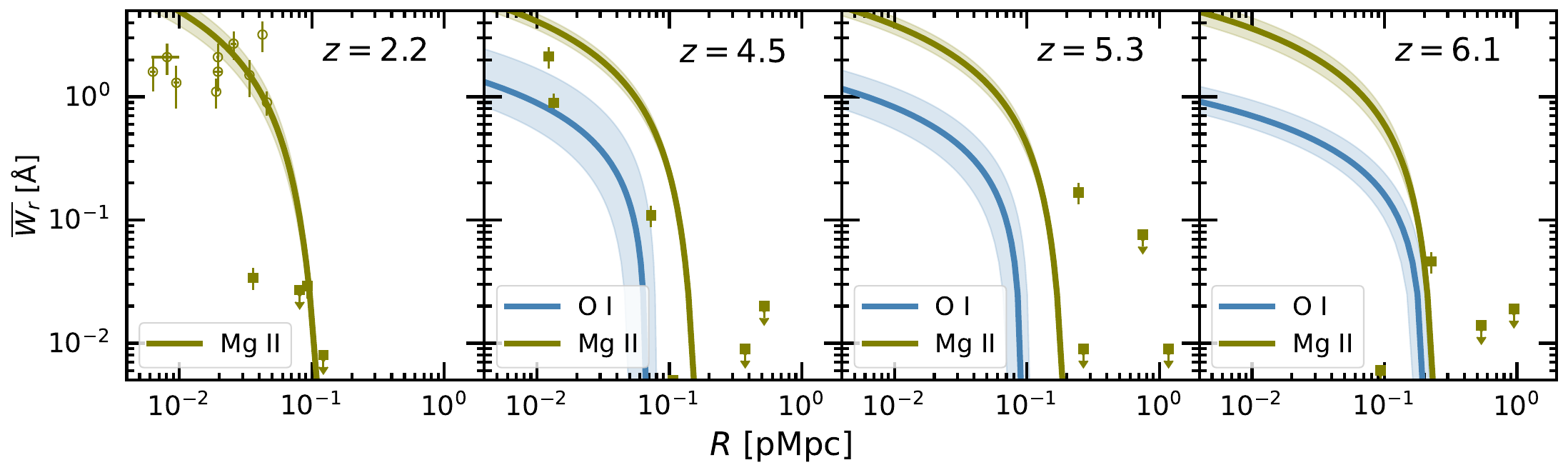}  
    \caption{Redshift snapshots of the predicted
      equivalent-width profiles, 
      $\Waver$, 
      of 
        \mgii\ $\lambda 2796$ and
    \oi\ $\lambda 1302$.  The redshifts are selected to
      match the available $W_r$-$\rho$ data at $z > 2$ for \mgii\ and
      the available $f(W)$ parametrization {  for
      \oi~\citep{Becker2019}.}
    {Model uncertainties are propagated from those of
      $f$ and $\phi$. } 
    The \mgii\ data points are taken from
    \citet{Bouche2012a} and \citet{Moller2020} (open circles),
    and~\citet{Bordoloi2024} (filled squares), with impact parameters normalized 
    by $(L/L^\star)^\beta$. 
    In the~\citet{Moller2020} data, $W_r$      
    is estimated based on the published velocity width,  
    by assuming a simple box profile of a fully saturated line.
  }
  \label{fig_profile_evolution}
\end{figure*}

\begin{table}
  \begin{center}
  \begin{footnotesize}
\caption{Best-fit results for redshift parametrization of Schechter
  parameters}
\label{table_fit}
\begin{tabular}{lcccc}
\hline
\hline
&$a$&$b$&$c$&$d$\\
$W^\star$&
1.24 $\pm$ 1.23 &
1.21 $\pm$ 2.93 &
2.75 $\pm$ 4.2 &
2.91 $\pm$ 1.25
\\
$N^\star$&
0.24 $\pm$ 0.04 &
1.11 $\pm$ 0.12 
\\
$\alpha$&
-1.32 $\pm$ 0.11 &
-0.37 $\pm$ 0.09 
\\
\hline
\end{tabular}
  \end{footnotesize}
  \end{center}
\end{table}

Building on the above procedure, we next investigated
\mgii\ profiles (and hence CGM cross sections) up to the highest
redshifts that absorption-line surveys allow.  For the UV-galaxy
number density, we again applied the parametrization of~\cite{Bouwens2021}, noting that while mergers are inherently considered
in $\phi(z)$, in the present approach we ignore their possible effect
on the absorbing cross sections~\citep{Hani2018}. We integrated
Eq.~\ref{eq_hits} from $L_{min}=0.01L^\star$, corresponding to
$M_B=-16$, i.e., the faintest magnitude bin of \citet{Bouwens2021} at
$z=6$, to $L_{max}=\infty$. To estimate the redshift evolution in
$f(z,W)$, we attempted to parametrize 
in redshift the 
Schechter parameters reported  
to fit the still-sparse data for \fz. At $z<2$, we used the Schechter
parameters reported by~\citet[][M17]{Mathes2017} for four redshift bins,
correcting their $\phi^\star$ to meet our definition of $N^\star$.
At $z>2$, to obtain the Schechter parameters, we first re-fit the
\fz\ data in the four redshift bins reported by~\citet[][C17]{Chen2017} using
a Schechter function (they originally fit an exponential
function).  We complemented their high-redshift bin with the measurement
reported by~\citet[][B17]{Bosman2017}.  We used a Bayesian approach to
explore correlations between parameters and constrain $N^\star$ such
that the area below the Schechter curve equals the total \dndz\ of the
survey.  The results, shown in Fig.~\ref{fig_schechter}, are
encouraging and provide robust constraints on $W^\star$ and $\alpha$,
despite the limited number of data points per redshift bin. With these eight sets of
measured Schechter parameters, plus the only other one reported at
$z>2$~\citep{Sebastian2024}, we explored the redshift evolution using
different parametrizations, as shown in
Fig.~\ref{fig_f_params_schechter}.  For $W^\star$, we followed the
prescription of~\citet{Madau2014}, used to fit
the evolution of 
the SFR density.  
For
$N^\star$ and $\alpha$, we fit simple power laws.  These are
functional fits without any particular physical motivation. However,
the~\citet{Madau2014} parametrization has been used to describe the
$d{\rm N}/{\rm d}z(z,W_r>1 {\rm \AA})$ data~\citep{Zhu2013}, which is
essentially determined by $W^\star$ (both in their exponential
parametrization and in our Schechter one).  The \dndz\ peak observed
at $z\approx 2$ (here seen in $W^\star$; upper panel of the figure)
has already been discussed by various authors and is believed to relate to the peak of cosmic star formation.  In fact, the UV
luminosity density, needed to obtain the SFR density~\citep[e.g.,
][]{Khusanova2020}, is proportional to the right-hand side of
Eq.~\ref{eq_hits} if $\beta=0.5$; in this case, the proportionality
factor is simply $L^\star$. This suggests that $W^\star$ should
continue to decrease with redshift and that the increase in $\alpha$ is
consistent with the anticorrelation between these two parameters, as observed in Fig.~\ref{fig_schechter}.  At the same time,
$N^\star$ defines the integral of $f(W)$ and balances the apparent
constant line density at high redshift. It remains to be determined
whether $N^\star$ continues to rise beyond $z\approx 6$, as
forced by the simple function used here, {  but we argue in
  \S~\ref{sect_discussion} that this is unlikely.}

\section{Results}
\label{sect_results}

With the redshift parametrizations of $f(z,W)$ we can now make
various statistical predictions. First, in
Fig.~\ref{fig_profile_evolution}, we show redshift snapshots of the
\mgii\ radial profile ($\kappa=1$). The redshifts are selected for
comparison with the few available $W_r$-$\rho$ measurements at $z>2$
\citep{Bouche2012a, Moller2020, Bordoloi2024}, and with
three redshifts with \fz\ data on neutral oxygen~\citep{Becker2019},
for which we can apply our formalism.  The \mgii\ data points are
moderately consistent with our statistical prediction, but more data are
needed to make a similar comparison to the low-$z$ one in
\S~\ref{sect_lowz}.  { Unfortunately, there are currently no
  observational data on $W_r^{1302}$ versus $\rho$ to independently
  validate our model on \oi.}  { However, a comparison between model
  predictions for \oi\ and \mgii\ is worthwhile. } In self-shielded neutral gas, \oi\ closely follows \hi\ ~\citep[e.g.,
][]{Keating2014} and has been proposed as a tracer of the \hi\ neutral fraction~\citep{Becker2019,Doughty2019}. In contrast, \mgii\ can
occur in both neutral and ionized gas.  Therefore, a comparison of the
radial profiles of these two species offers good prospects for tracing
the growth of metal-enriched bubbles in the pre-reionized IGM.  {
  Focusing on $W_r^{2796}>0.3$~\AA\ \mgii\ and
  $W_r^{1302}>0.1$~\AA\ \oi\ (where $f(W)$ is still complete)}, just
after reionization ($z\sim6$), our statistical comparison shows that
the two ions trace each other similarly, whereas at lower redshifts
\oi\ is less extended than \mgii. This is most likely due to
ionization effects, since \mgii\ can survive under slightly higher
ionization conditions~\citep[and not due to evolution in enrichment of
  the CGM; ][]{Doughty2019}.  We analyze this scenario in more detail
in~\S~\ref{sect_discussion}.

\begin{figure}
  \centering
  \includegraphics[width=1\columnwidth]{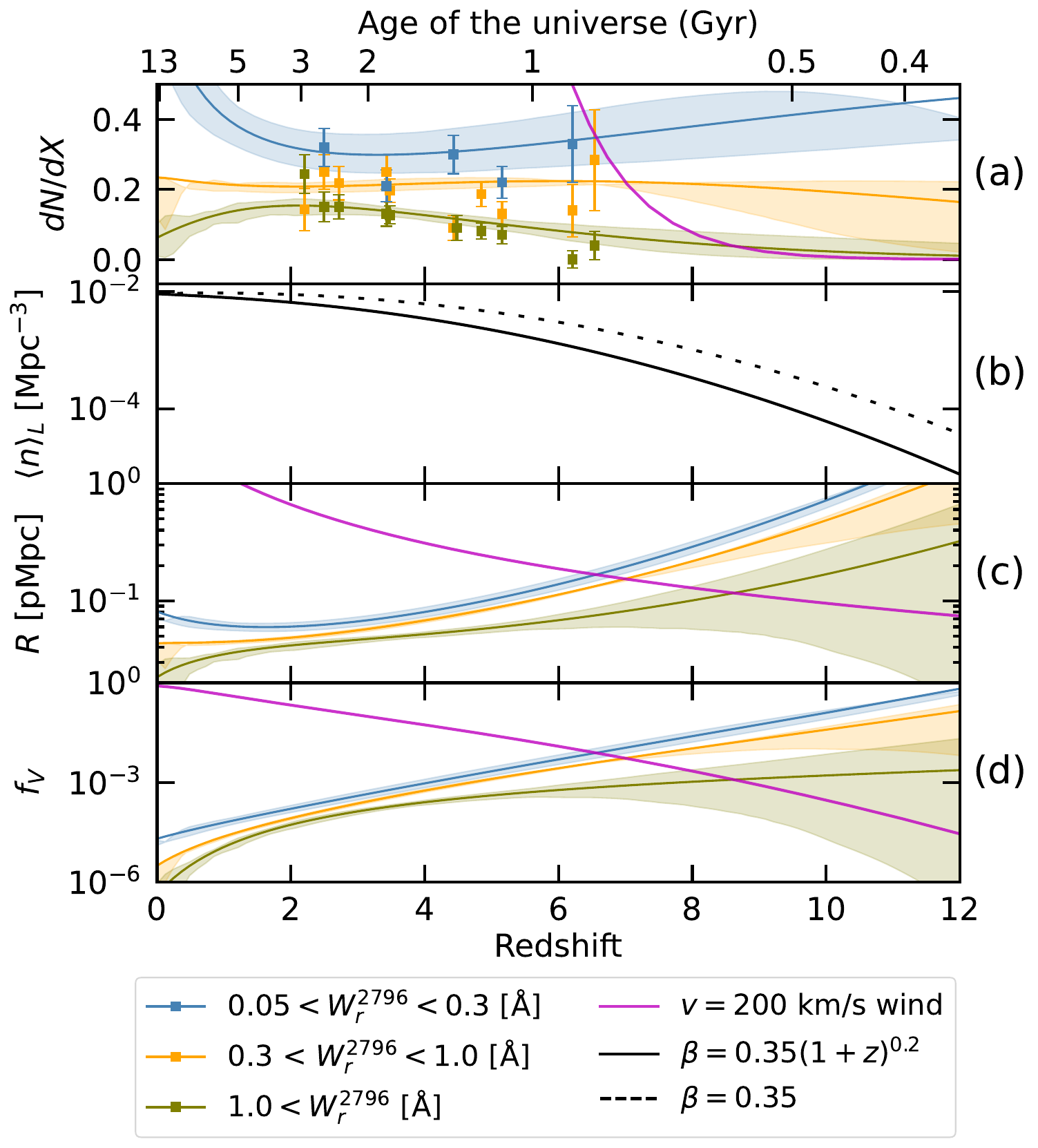}
  \caption{ Redshift evolution of various
    model 
    quantities,   
    shown as following from top to
    bottom.  
    Panel (a): Statistical prediction of the \mgii\ comoving line
    density \dndx\ for three cuts in $W_r^{2796}$ {  chosen to match
    published data} (colored lines). 
    Data points with the same color codes are 
    from~\citet{Sebastian2024} and \citet{Chen2017}. 
   Panel (b): Observed
    $L>0.01L^\star$ luminosity-weighted space density of UV-bright 
    galaxies~\citep{Bouwens2021}  shown for both fixed and evolving $\beta$. Panel (c):
    Statistical prediction of the 
    \mgii\ halo radius. Panel (d): Same as panel (c) but 
    for the volume filling factor.
    {  Since $W_r$ is binned, both $R$ and  $f_V$ are averages. 
 The magenta line in panels (a), (c), and (d) represents a
 constant-velocity wind starting at the Big Bang. }
        {  Model uncertainties are propagated from those
          of       $f$ and $\phi$.  
    A non-evolving $\beta$ would shift $\langle n \rangle_L$ upward by a factor of $\approx 3$, and shift $R$ and $f_V$ downward by  $\approx 1.7$ and $\approx 5$, respectively.} 
  }
  \label{fig_evolution}
\end{figure}

A more general evolutionary view is shown in Fig.~\ref{fig_evolution},
which displays for 
our \mgii\ model:  
 (a) the predicted redshift evolution of the
comoving line density\footnote{ \dndx\ is obtained by integrating
  Eq.~\ref{eq_f} and dividing by d$X/$d$z$ to remove the cosmological
  effect on the line density.}, \dndx, in three $W_r$ intervals; (b)
the luminosity-weighted space density of galaxies, $\langle
n\rangle_L$; (c) the radius, $R$, (for $\kappa=1$); and (d) the volume
filling factor, defined here as
\begin{equation}
f_V \equiv \frac{4\pi}{3}R^3~n_c~(1+z)^3~.
\end{equation}
In panel (a), {  the selected $W_r$ intervals enable comparison with the
  binned data reported in the literature, 
  i.e.,~\citet{Chen2017}  
and~\citet{Sebastian2024}.} Since these   
constrain the \fz\ {  data} used in~\S~\ref{sect_implementation},  
it is not surprising that they 
capture the predicted evolution in \dndx.  From
$z\approx 6$ to $z\approx 2$, the line density of 
$W^{2796}_r>1$ \AA\ absorbers steadily increases with time by a factor of
{  $\sim 
  3$}, as noted 
in~\citep{Matejek2012,Zhu2013,Chen2017}. This is in agreement with
our discussion above on $W^\star(z)$, since \dndx($W_r>1$ \AA) is
dominated by $W^\star$.  In contrast, the line density of weak
absorbers ($W_r <0.3$ \AA) decreases {  with time} in the same
{  redshift}
interval.  We speculate that this could be due to the complete
transmission of ionizing photons upon reionization, which affects the
outermost layers of the model (i.e., weak systems) more strongly than
the more self-shielded inner layers (i.e., strong systems).
The onset of a hard ionizing UV background (UVB) has been linked to
the evolution of the ratio 
of low to high ions upon
reionization~\citep{Finlator2016,Becker2019,Cooper2019,DOdorico2022},
although radiation-driven outflows can also remove gas   from the   
 low-ionization phase, and these may be more common in the early
 Universe~\citep{Ferrara2024}.   
 Together with a steady increase in the galaxy space density with time,
{  panel (c) shows how} these effects are reflected in a $2.5$-fold 
decrease in the 
radial extent of weak \mgii\ gas, but only a $1.5$-fold for 
very strong
\mgii\ gas. Hence, the end of
reionization could be largely governed by a balance between the growth
of metal-enriched, supernova-driven bubbles and regions of photoionized hydrogen, with the former growing more slowly than the latter.

At $z\approx 6$, the average proper distance between $L>0.01L^\star$
galaxies is $\sim 2$ Mpc, a factor of $\approx 10$ larger than the
\mgii\ radius predicted here for the unity covering fraction,
$\kappa=1$. Winds apparently have not been able to reach neighboring
halos.  This is reflected in the predicted low filling factor,
$f_V\lesssim 1$ \%, and is consistent with
theory~\citep[e.g.,,][]{Tie2022,Madau2001}.  For the enriched halos to
have a linear extent ten times larger, and {  therefore be able to
  reach their neighbors}, $\kappa\approx 0.01$ would be required
(Eq.~\ref{eq_kappa}). {  In this case, however, the cross section for
  absorption would be very low~\citep[much lower than observed at
    $z\sim 1$; e.g., ][]{Schroetter2021}}, and consequently the
``crossing probability'' would also be low. We therefore
conclude that, at the end of reionization, the winds have not yet
reached the neighboring halos. This implies that halos have so far
been mostly self-enriching, as suggested by semi-analytical
models~\citep{Ventura2024}, in a process mixed with the steady
replenishment of IGM material~\citep{Waterval2025}.

{  Our results depend on the chosen parametrization of $f(z,W)$, and
  in \S~\ref{sect_discussion} we argue that it cannot be extrapolated
  much beyond $z\approx6$--$8$.  The uncertainty of the model is
  dominated by that in the $f(z,W)$ parametrization.  Regarding the
  halo-size-luminosity scaling, an unlikely non-evolving $\beta=0.35$
  shifts $\langle n \rangle_L$ upward by a factor of $\approx 3$ (without
  affecting the comparison between weak and strong absorbers), and $R$
  and $f_V$ downward by factors of $\approx 1.7$ and $\approx 5$,
  respectively, which strengthens the prediction of a low filling
  factor.  }

\section{Discussion}
\label{sect_discussion}

Within and beyond the reionization epoch {  in redshift}, this regime remains largely uncharted. 
According to our prediction, the average metal
extent beyond $z=6$ {  continuously increases with
  redshift}. However, this contradicts early chemical enrichment
models, in which metal-enriched bubbles {  expand over time}~\citep[e.g.,
][]{Furlanetto2003,Wang2012,Finlator2020,Yamaguchi2023}. In
Fig.~\ref{fig_evolution} panels (a), (c) and (d) the magenta lines 
display 
a toy model in which constant-velocity isotropic winds have expanded
around all halos since the Big Bang. 
Here we use
$v=200$ \kms, which is well above the maximum velocity for any
dark-matter halo mass in wind models~\citep[e.g.,
][]{Furlanetto2003,Yamaguchi2023}.  {  Thus, this simple wind model
  sets a firm upper bound for the extent of metals before
  reionization is complete. It also suggests that a high-redshift
  extrapolation of our predictions does not hold and that a sharp decline
  in \dndx\ should be observed beyond $z=6$--$8$.}

{  Analytical models~\citep[e.g., ][]{Furlanetto2003} and
  hydrodynamical simulations~\citep[e.g.,
  ][]{Oppenheimer2008,Keating2016,Finlator2020,Ocvirk2020} predict
  that the sizes of the metal-enriched regions (a few pkpc) and thus their volume filling factors, are significantly smaller than
  those of ionized bubbles (hundreds of pkpc). This results from
  differences in the mechanisms driving their
  expansion~\citep[mechanical feedback by winds and supernovae and stalled
    expansion due to metal cooling in the first case, versus radiative
    feedback in the second; e.g., ][]{Madau2001}.  As a result, the
  metal-enriched bubbles expand in an already ionized
  medium~\citep[e.g., ][]{Shin2008} and have a complex, stratified
  ionization structure~\citep{Oppenheimer2009,Furlanetto2005}.  This
  complexity has made it difficult for simulations to simultaneously
  reproduce \dndx\ for species probing different ionization states.

  Our model predicts an evolution in redshift toward more similar
  profiles of strong \mgii\ and
  \oi\ (Fig.~\ref{fig_profile_evolution}), whereas a simple expanding
  wind model suggests a decreasing line density for both ions.  The
  question arises as to whether these two predictions are consistent
  with each other.  The first prediction is consistent with covering
  fraction arguments. Such high $W_r$ values are only reached if
  multiple optically thick clouds are intercepted by the line of
  sight. This is corroborated by the large number of absorption
  components typically observed in such high-$W_r$ systems and by
  numerical simulations, in which neutral gas clouds in expanding
  outflows have sizes of at most a few hundred
  parsecs~\citep[e.g.,][]{Dutta2025}, small compared to the extent of
  the winds, which reach at least kiloparsec scales
  \citep{Yamaguchi2023, Finlator2020}. Therefore, the absorption cross
  section, and thus \dndx, is dominated by the projected extent over
  which multiple optically thick clouds are present in an otherwise
  ionized volume-filling phase.  Each cloud may have a partially
  ionized outer layer containing only \mgii\ and no \oi\ (analogous to
  classical photodissociation regions); however, in a swarm of clouds
  with high covering fraction, this layer will not dominate the
  \mgii+\oi\ cross section, leading to similar radial profiles for
  both ions.  By the end of reionization, the rising ionizing
  background~\citep[e.g., ][]{Davies2024} fundamentally alters this
  scenario. Neutral clouds in the outer parts of the halo become
  depleted in \oi, whereas \mgii\ can survive more easily. This change
  in the source of ionizing photons, coupled with the dramatic
  increase in halo number density, is consistent with the second
  prediction, and together both factors support the evolutionary
  trends shown in Figures~\ref{fig_profile_evolution}
  and~\ref{fig_evolution}.  }

\section{Conclusions}
\label{sect_conclusions}

Our considerations, based on observational data, assumptions about the
halo geometry, and theoretical models of early metal
enrichment, predict a complete evolutionary pathway for low-ionization
metal-enriched halos. Specifically, we observe a marked difference
between the strong and weak \mgii\ systems. At any redshift upon reionization, strong systems have consistently exhibited a lower incidence (and
therefore also a smaller cross section) than weak systems.  However,
both types of system grow during the first few Myr and peak between $z\approx 8$ and
$6$.  After reionization is complete, the incidence of strong systems,
coupled with steadily increasing star formation, increases, reaching
a peak at cosmic noon. The incidence of weak systems, because they are more exposed to the ambient UVB, steadily
decreases or remains constant until it turns over at cosmic noon.
By the end of reionization, the metals have not yet reached the
neighboring halos. Furthermore, the increase in halo number density
causes the average linear extent of both the strong and weak
\mgii\ halos to decrease over time, with the latter decreasing more
rapidly.  In our model, these evolutionary paths result from the
competing effects of increasing halo number density, expanding
metal-rich bubbles, and changing ionization conditions.

We propose that the present approach can continue to be used as new
high-redshift campaigns are implemented~\citep[e.g.,
][]{Bordoloi2024}.  Galaxy and absorption-line surveys are
established techniques with well-known selection functions. { Quasars
  and gamma-ray bursts are already being identified beyond $z\approx 8$ and will
  soon be within reach of JWST and ELT medium-resolution
  spectroscopy.}  The identification of intrinsically strong metal
transitions, unaffected by the Gunn-Peterson trough, appears feasible
down to weak limits in the bright continua of such sources.

\begin{acknowledgements}
  We thank the anonymous referee for their valuable criticisms.  This
  work has benefited greatly from conversations with
  Hsiao-Wen Chen, Lise Christensen,  
  Valentina D'Odorico,   
  and Claudio Lopez-Fernandez.
  S.L. acknowledges support by FONDECYT grant 1231187.

\end{acknowledgements}

\bibliographystyle{aa}
\bibliography{Lopez_lit}

\begin{appendix}
\section{Schechter function fits}
\begin{figure}[h!]
  \begin{center}
  \includegraphics[width=1\columnwidth]{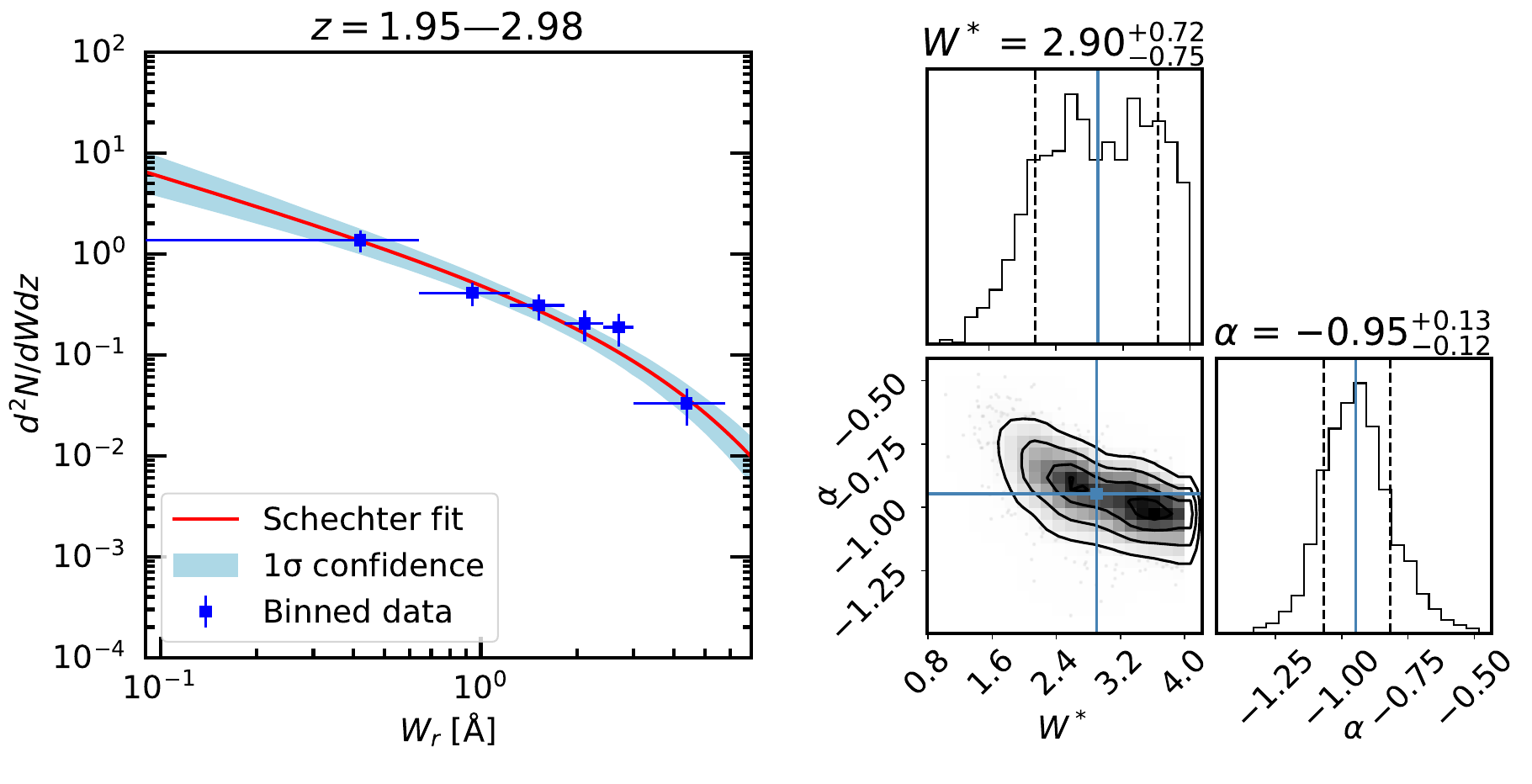}
  \includegraphics[width=1\columnwidth]{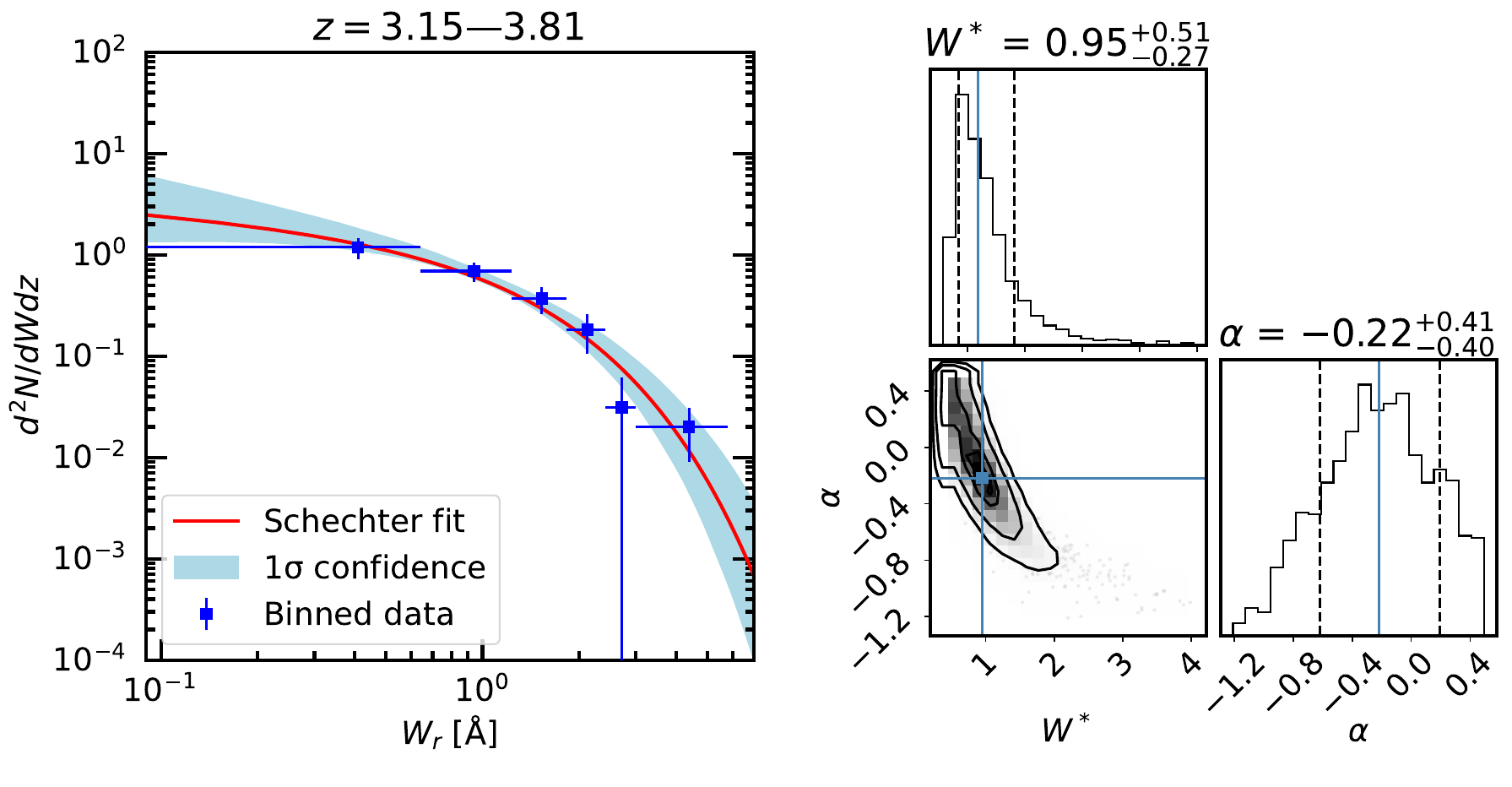}
  \includegraphics[width=1\columnwidth]{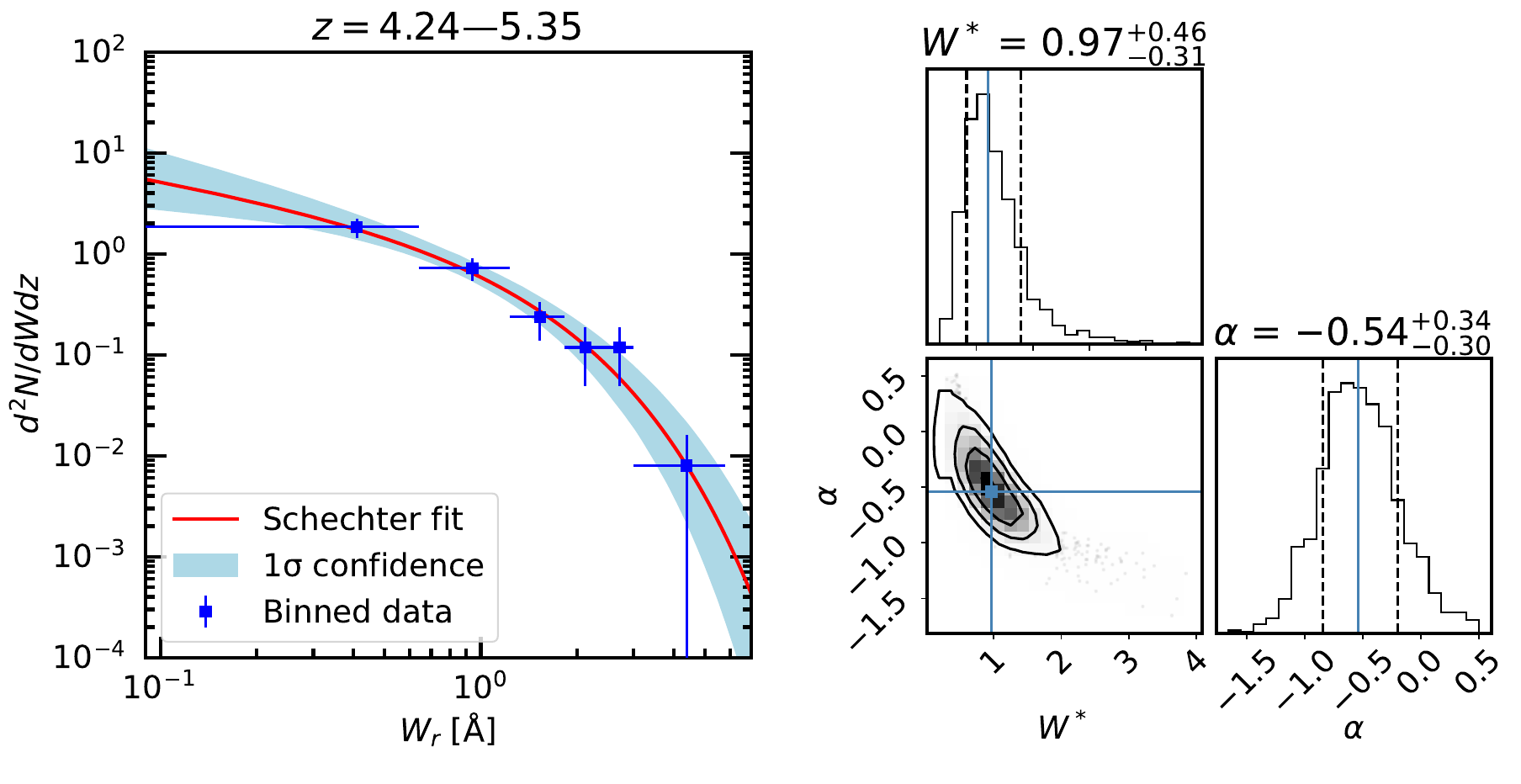}
  \includegraphics[width=1\columnwidth]{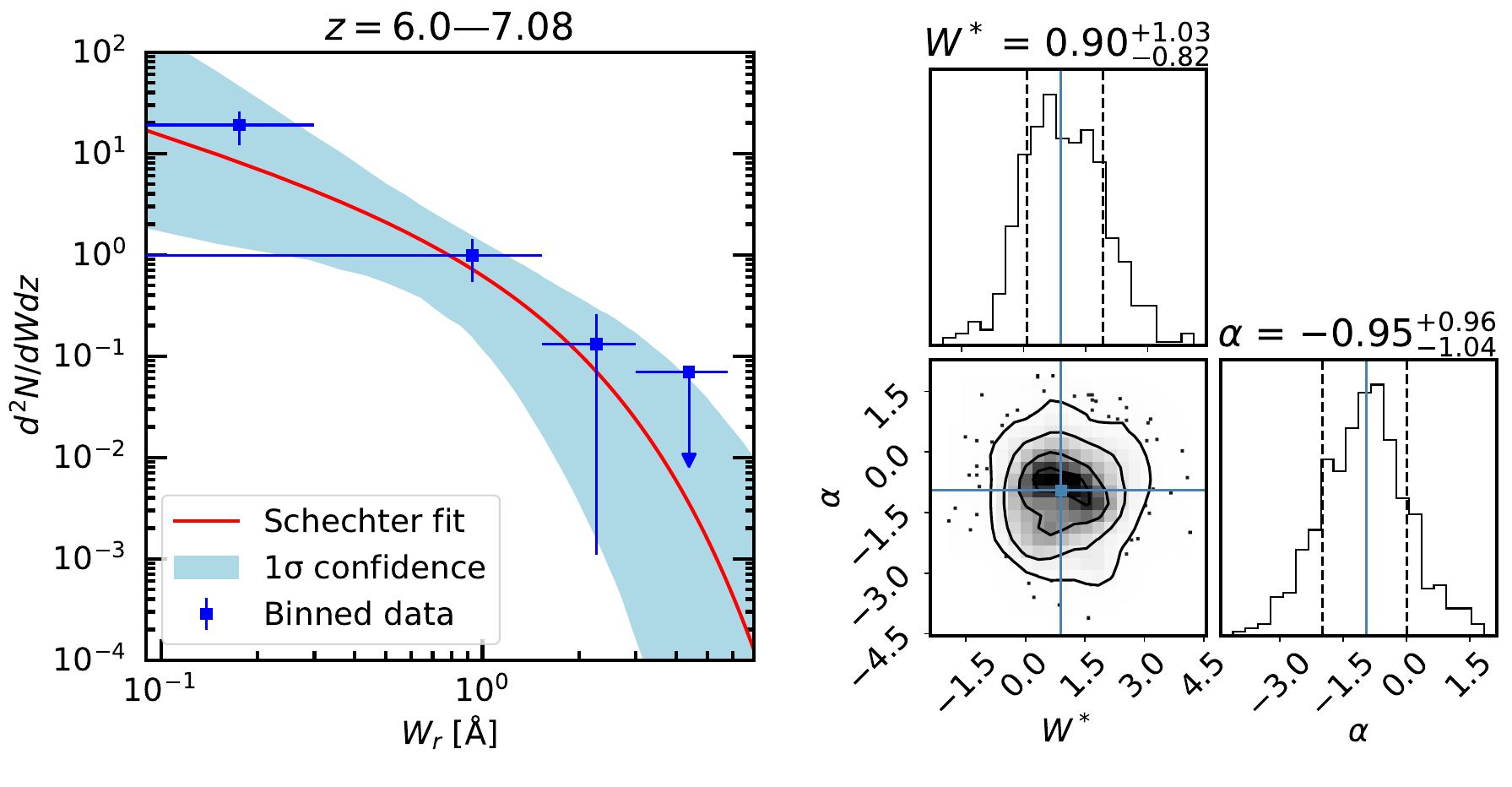}
  \end{center}
  \caption{ Schechter function Markov chain Monte Carlo (MCMC) fits to
    the \fz\ data by~\citet[][C17]{Chen2017}.  The $6.0<z<7.1$ bin is
    complemented with the low-$W_r$ measurement reported
    by~\citet{Bosman2017}.  The C17 upper limit is treated as a
    uniform prior truncated at the upper limit value.  The fits are
    constrained such that the area below the fit curve equals
    the total ${\rm d}N/{\rm d}z$ in the exponential fits, i.e., $N^
    \star$ in C17. This results in some correlation between the
    Schechter parameters $W^\star$ and $\alpha$. A pure exponential ($\alpha=0$) is ruled out at the $1$-$\sigma$ level in three cases. The $\pm$ 1-$\sigma$ confidence band around the fit curve
    is built from the MCMC samples.  }
  \label{fig_schechter}
\end{figure}

\end{appendix}
\end{document}